# Exploiting Multimodal Biometrics in E-Privacy Scheme for Electronic Health Records

Adebayo Omotosho[1,2*]     Omotanwa Adegbola[1]     Barakat Adelakin[1]     Adeyemi Adelakun[1]
Justice Emuoyibofarhe[2]
1.Department of Computer Science, Bells University of Technology, Ogun State, Nigeria
2.Department of Computer Science and Engineering Ladoke Akintola University of Technology, Oyo State Nigeria
* E-mail of the corresponding author: bayotosho@gmail.com

**Abstract**
Existing approaches to protect the privacy of Electronic Health Records (EHR) are either insufficient for existing medical laws or they are too restrictive in their usage. For example, smartcard-based encryption systems require the patient to be always present to authorize access to medical records. A major issue in EHR is how patient's privacy and confidentiality can be maintained because there are known scenarios where patients' health data have been abused and misused by those seeking to gain selfish interest from it. Another issue in EHR is how to provide adequate treatment and have access to the necessary information especially in pre-hospital care settings. Questionnaires were administered by 50 medical practitioners to identify and categorize different EHR attributes. The system was implemented using multimodal biometrics (fingerprint and iris) of patients to access patient record in pre-hospital care. The software development tools employed were JAVA and MySQL database. The system provides applicable security when patients' records are shared either with other practitioners, employers, organizations or research institutes. The result of the system evaluation shows that the average response time of 6seconds and 11.1 seconds for fingerprint and iris respectively after ten different simulations. **T**he system protects privacy and confidentiality by limiting the amount of data exposed to users. The system also enables emergency medical technicians to gain easy and reliable access to necessary attributes of patients' EHR while still maintaining the privacy and confidentiality of the data using the patient's fingerprint and iris.

## 1. Introduction
A patient's medical record accumulates significant personal information including identification, history of medical diagnosis, digital renderings of medical images, treatment received, medication history, dietary habits, genetic information, psychological profiles, employment history, income, and physicians' subjective assessments of personality and mental state among others (Mercuri, 2004). It may also contain patient report such as HIV/AIDS status, mental history, and pregnancy reports, some of this information need to be treated with the highest level of confidentiality.

Medical records are often vulnerable to abuse by those seeking to gain from it. There are good reasons for keeping the records private and limiting the access to only minimum necessary information for example an employer may decide not to hire someone with past psychological issues, an insurance company may refuse to provide life insurance when knowing the disease history of a patient, a person with certain types of disease may be discriminated by the healthcare provider, and so on (Sun *et al,* 2008). Also, first responders and emergency care providers make life or death decisions with little to no information about patient's medical problems therefore having sufficient information about the patient medical history would be helpful in saving the patient's life (Aguinaga and Poellabauer, 2012).

Some researchers also identified that emergency access represents one of the easiest methods to access unauthorized data because the hacker needs only to provide a plausible reason for access Darnasser (2013). Therefore, in this paper, we aim to identify and implement some essential attributes of patient's health record that should be made available at all times and shared, and attributes that should be revealed only during emergency situation or pre-hospital care. Creating emergency attributes from patients' health record to be accessed in emergency situation is an additional measure that enables technicians to gain access to necessary data in pre-hospital care while preserving patient's privacy and confidentiality.

## 2. Related Work
In many countries privacy laws are required to protect the confidentiality of EHR records and let the patient control the access to them. Existing approaches to protect the privacy of EHRs are either insufficient for these laws or they are too restrictive in their usage. For example, smartcard-based encryption systems require the patient to be always present to authorize access to medical records. However, this does not allow a physician to access an EHR of a patient who is unable to show up in person. Hupperich *et al*, (2003), proposed security architecture for EHR infrastructure that provides flexibility. The security of this approach relies on modern cryptography schemes and their incorporation into an EHR infrastructure. The model enables patients to authorize access to their records remotely (e.g. via phone) and time-independent for later processing by the physician. However, in emergency cases, patient might be unconscious or otherwise not able to authorize access to health record.

Krawezyk *et a*l, (2005), analyzed the performance of combing the use of on-line signature and voice biometrics in order to perform user authentication in securing electronic medical record. This paper however, did not consider patients during emergency situations that are unable to speak. The use of on-line signature biometrics also poses as a disadvantage in this work because of its inherent high intra-class variability.  Appari *et al*, (2008) critically analyzed information security and privacy in healthcare, and also developed technological solutions for ensuring privacy of patients when their information is stored, processed and shared. This research failed to show how privacy will be maintained at all levels. Similarly, Mohan *et al*, (2008), implemented a framework for electronic health record sharing that is patient centric. This approach demonstrated a scenario involving emergency responders' access to health record information but focused mainly on internal threat and EHR sharing.

Narayan *et al*, (2010), showed how to provide privacy in e-health systems with attribute-based encryption. However, their approach is based on a specific structure for health records and in their model patients administers their own health records as PHRs. The authors do not discuss transferrable authorization secrets, but let patients define encryption policies .They assume that patients know whom to authorize when they create (or re-encrypt) EHRs. Aguinaga and Poellabauer (2012) used Quick Response (QR) codes and smart phones to access emergency information while preserving patient privacy of sensitive information. The method used was to embed emergency information on smart phone's lock screen in the form of a two-dimensional barcode. The code is then scanned and decoded using the same smart phone app or any QR code scanner. This is dependent on the assumption that a patient will be carrying the smart phone during emergency, thus it is considered to be impractical for the basis of health care.

Benaloh *et al*, (2011), this considered an efficient system that allows patient to share partial access right with others and to operate searches over their records. The drawbacks of their proposed solution however are the need to create and manage multiple keys by patients and healthcare providers, the absence of an efficient user revocation mechanism, need for an external key escrow agent, and the need for patients to verify the healthcare provider's credentials.

Sun *et al,* (2011), proposes a secured EHR system, HealthCare system for Patient Privacy (HCPP) based on cryptographic constructions and existing wireless network. Their work illustrates design protocols for a secure healthcare system leveraging cryptographic tools. Darnasser, (2013) focus on separating emergency data from the core of EHR systems in order to minimize the amount of leaked data in cases of emergency. The model made use of multiple cloud providers whereby data is being divided between those providers and the system only knows what data is stored in each and used "push" and "pull" approach to develop a generalized method for enhancing patient privacy during emergency access.

Recently, Diaz *et al*, (2013), propose the use of biometric identification to access a central health record database. The method used was to provide the technicians with a mobile system through which they gain access to necessary attributes of patients EHR using the patients fingerprint during emergency. Although, this paper focuses on granting proper access to a patient's health record remotely with the use of biometric identification system. It uses only patient's fingerprint for authentication (unimodal biometrics), which is considered to be a weak implementation because unimodal biometrics may suffer from the following problems listed below. Also, the experiment demonstrated an average response time of 19.8 seconds with about two hundred thousand patients registered in the database. In this paper, we aim to improve on this by exploiting multimodal biometrics and comparing it with a larger population and aim to get a faster response time.

Some problems associated with unimodal biometrics are include
- Noise in sensed data -A fingerprint image with a scar or a voice sample altered by cold is an example of noisy data.
- Intra-class variations - These variations are typically caused by a user who is incorrectly interacting with the sensor (e.g., incorrect facial pose), or when the characteristics of a sensor are modified during authentication (e.g., optical versus solid-state fingerprint sensors).
- Inter-class similarities- In a biometric system comprising of a large number of users, there may be inter-class similarities (overlap) in the feature space of multiple users..
- Non-universality- The biometric system may not be able to acquire meaningful biometric data from a subset of users. A fingerprint biometric system, for example, may extract incorrect minutiae features from the fingerprints of certain individuals, due to the poor quality of the ridges. Multimodal addresses the problem of non-universality, since multiple traits ensure sufficient population coverage.
- Spoof attacks - This type of attack is especially relevant when behavioral traits such as signature or voice are used. However, physical traits such as fingerprints are also susceptible to spoof attacks.

Multimodal also deters spoofing since it would be difficult for an impostor to spoof multiple biometric traits of a genuine user simultaneously.

## 3. Materials and Methods
*3.1 Data Collection*

Questionnaires were designed after the initial consultations with medical practitioner to acquire information on how to classify health data into categories such as: Basic, Confidential and Emergency attributes with respect to record sharing. The questionnaires were filled by medical practitioners in Lagos State University Teaching Hospital (LASUTH), Ogun State Hospital, Ota and some other private hospitals within Ota and Lagos metropolis, Nigeria. The research questionnaire designed comprises headings like Patients Bio data, Common Medical Conditions and History, Psychiatric Medical Conditions and History, Common Surgical Conditions and History and Statuses. The questionnaire aim is to identifying the basic, confidential and emergency attributes of patients' health data under the various headings. With this we were able to achieve a goal that provides applicable privacy when patients' records are shared either with other practitioners, employers, organizations or research institutes. Also, the system was able to identify the necessary information needed during emergency situations and allow emergency technicians (paramedics) to have access to these data through patients fingerprint or iris.

The research questionnaire was obtained from seventeen different hospitals. The average years of practice of the medical practitioners that filled the questionnaires were approximately eight years.

Table 1, Table 2 and Table 3 depicts the summary of the EHR data sharing rule that was formulated from the result of the fifty questionnaires showing the identified attributes categorized under Basic, Confidential and Emergency. The proposed rule nesting is depicted in Figure 1. Health records of patient that can be shared among various hospitals, organizations, employers and other institution or when revealed does not cause any harm to the patients are classified as Basic attributes. Patients' health records that should not be available at all times and cannot be shared without the patient permission are classified as Confidential Attributes. While, Emergency attributes are patient health data that should be available during emergency situations. Privacy of patient is preserved by enforcing the e-privacy rule because it helps to limit the number of shared or exposed data about the patient.

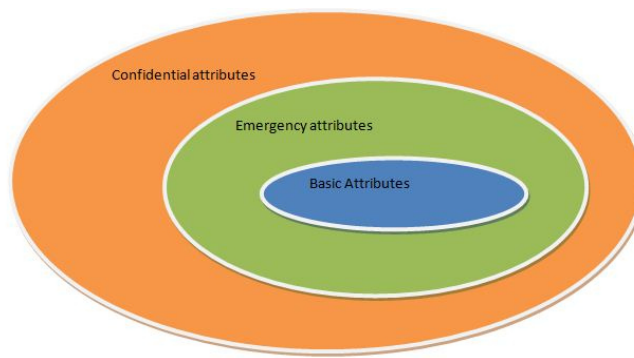

Figure 1 Attributes nesting

Table 1. Identified Basic Attributes

| Patients Bio Data | Common Medical Conditions | Psychiatric Medical Conditions | Common Surgical Conditions | Statuses |
|---|---|---|---|---|
| <ul><li>Name</li><li>Age</li><li>Sex</li><li>Religion</li><li>Nationality</li><li>Marital status</li><li>Parity</li></ul> | <ul><li>Hypertension</li><li>Diabetes</li><li>Dyslipidemia/Hypercholesterolemia</li><li>Arthritis</li><li>Arrhythmia</li><li>Chronic kidney disease</li><li>Cancer</li><li>Recurrent urinary tract infection</li><li>Chronic Obstructive pulmonary disease (COPD)</li><li>Medical implant</li><li>Asthma</li><li>Congestive heart failure</li><li>Myocardial Infarction/ Angina</li><li>Coronary Artery disease</li><li>Inflammatory bowel disease</li><li>Parkinson disease</li></ul> | <ul><li>Autism</li><li>Mania</li><li>Depressive illness</li></ul> | <ul><li>Past surgeries</li><li>Surgical implants</li></ul> | <ul><li>Blood group</li><li>Genotype</li></ul> |

Table 2 Identified Confidential Attributes

| Patients Bio Data | Common Medical Conditions | Psychiatric Medical Conditions | Common Surgical Conditions | Statuses |
|---|---|---|---|---|
| Sexuality | Epilepsy | <ul><li>Schizophrenia</li><li>mania</li></ul> | Past surgeries | <ul><li>HIV/AIDS</li><li>Hepatitis B</li><li>Hepatitis C</li></ul> |

Table 3 Identified Emergency Attributes

| Patients Bio Data | Common Medical Conditions | Psychiatric Medical Conditions | Common Surgical Conditions | Statuses |
|---|---|---|---|---|
| ➢ Age<br>➢ Parity | ➢ Hypertension<br>➢ Diabetes<br>➢ Dyslipidemia/ Hypercholesterolemia<br>➢ Arthritis<br>➢ Arrhythmia<br>➢ Chronic Kidney Disease<br>➢ Cancer (e.g. Leukemia, Brest Cancer)<br>➢ Recurrent Urinary Tract Infection<br>➢ Chronic Obstructive Pulmonary Disease (COPD)<br>➢ Medical Implants (Pacemaker)<br>➢ Asthma<br>➢ Congestive Heart Failure<br>➢ Epilepsy<br>➢ Parkinson's Disease | - | ➢ Benigh Prostatic Hyperplasia (BPH)<br>➢ Past Surgeries<br>➢ Surgical Implants | ➢ Blood Group<br>➢ Genotype<br>➢ HIV/AIDS<br>➢ Hepatitis B<br>➢ Hepatitis C |

*3.2 The Conceptual Architecture*

Figure 2 shows the proposed system architecture in which mulitimodal biometrics is used in forming e-privacy rule. The architecture shows the different components that would make up the system which includes the patient, healthcare provider or paramedics, the EHR database and other agencies. Each of these components performs its own functions and they interact to form the system as a whole. The system architecture describes the process in which the paramedics or health provider gain adequate access to the patient's record through patient's biometric authentication (fingerprint and iris). It depicts a situation whereby a paramedic gets to an unconscious patient who needs medical attention; the paramedics gain access in to the patient's health record through the patients fingerprint and iris thereby, emergency attributes which was gotten from the analysis of the questionnaires distributed are revealed. These *Emergency attributes* contains information that should be sufficient enough for paramedics to work with.

    **The paramedic** is the individual who attends to the unconscious patient in pre hospital care. The paramedic is provided with a mobile device through which he gains access to the *necessary attributes* using patients fingerprint and iris.

    **The EHR database** is where all patients' health records are stored which includes the basic, confidential and emergency attributes. The EHR database also retrieves and filters patient's record.

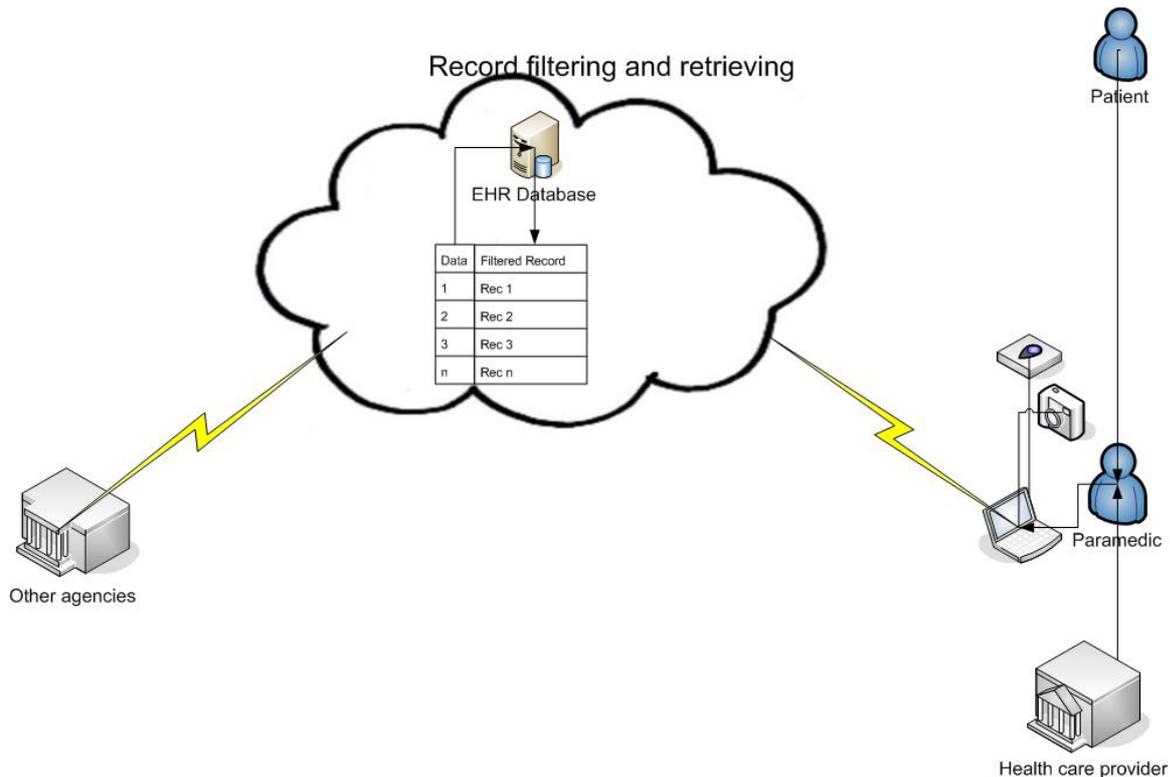

*Figure 2 Conceptual Model*

## 4. Results and Discussion
*4.1 Program Module And Interface*
The EHR system that implemented the designed rule was designed using JAVA programming and MySQL database. The interfaces are shown below:

*4.1.1 Authentication Interface*
Figure 3 represents the authentication interface, this is the first screen shown when the application is started. It allows the user specify the method of authentication which is either with iris or fingerprint. Doctors can also Login through this interface.

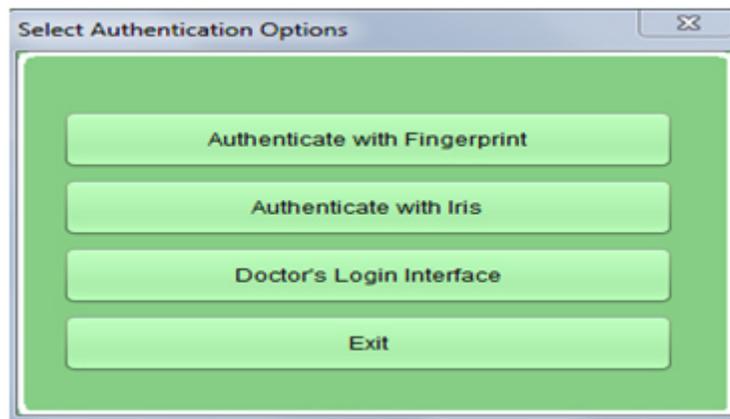

Figure 3 Authentication Interface

*4.1.2 Fingerprint Authentication Screen*
Figure 4 represents the fingerprint authentication whereby a user (paramedics) might use the patients fingerprint to access patient health record.

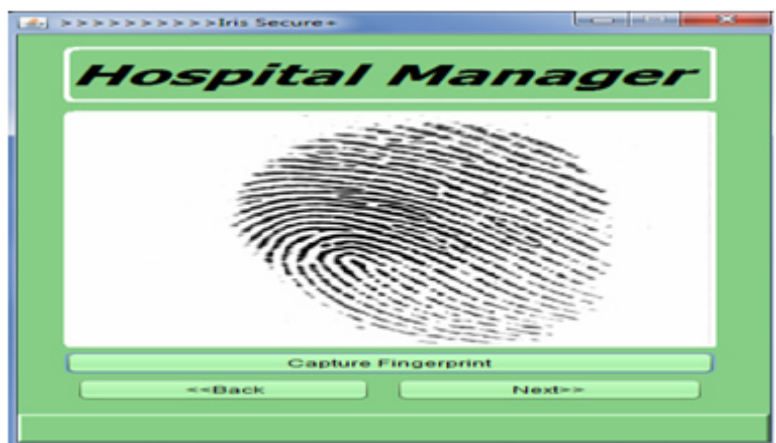

Figure 4 Fingerprint Authentication Screen

*4.1.3 Iris Authentication Screen*

Figure 5 represents the alternative biometrics, iris authentication, whereby a user (paramedics) might use the patient iris to access patient's health record.

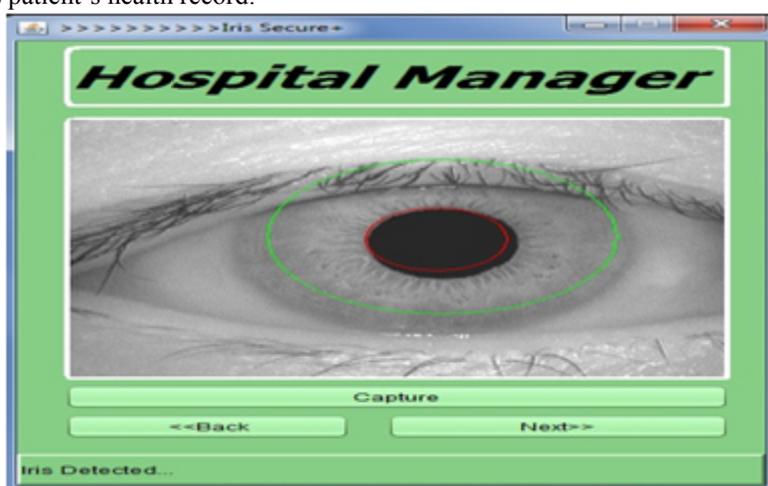

Figure 5 Iris Authentication Screen

*4.1.4 User Option after the Authentication Screen*

After a successful biometric authentication a doctor or paramedic is provided with three options as shown in Figure 6. Users can register patient, open emergency EHR or update (populate) a new created electronic record data for newly registered patient.

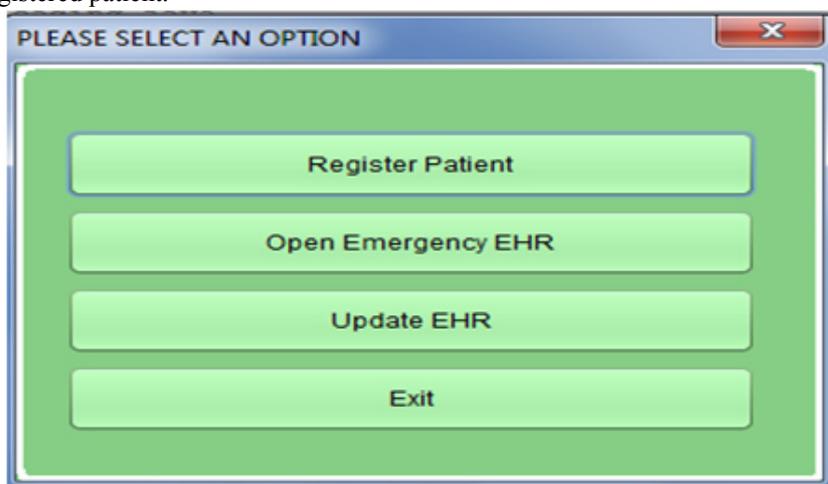

Figure 6 User Option after the Authentication Screen

*4.1.5 Registration Page*
Figure 7 shows the patient registration page for a new patient. It provides details about the patient

Figure 7 Registration page

*4.1.6 Emergency Attributes in EHR*
Figure 8 shows only the emergency data for a sample patient, displaying only the records filtered based on the sharing table arrived at from the questionnaire.

Figure 8 Emergency EHR

*4.1.7 Login Page for Medical Doctors*
This is the general login interface where doctors can access their patients records based on their access level.

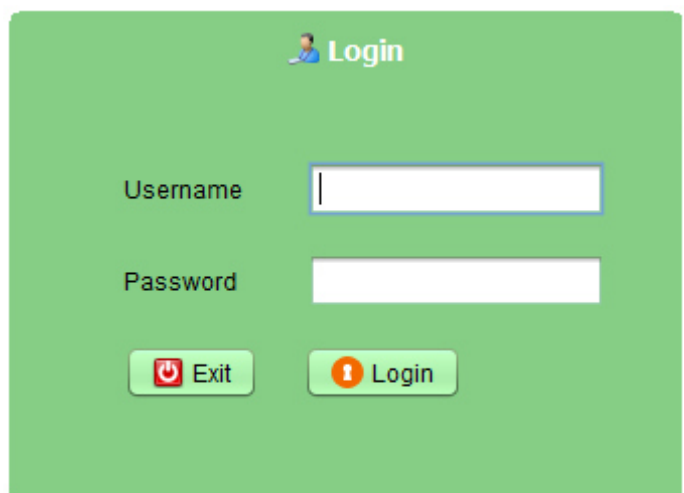

Figure 9 Login Page for Medical Doctor

*4.1.8 Medical Doctor's Interface*
Figure 10 represents the doctor's interface whereby medical doctors can manage patients and access patient basic health record through the patient reference number. Doctors can also view patient confidential attributes by using the patient private key. It also shows the list of registered patient and their popup options.

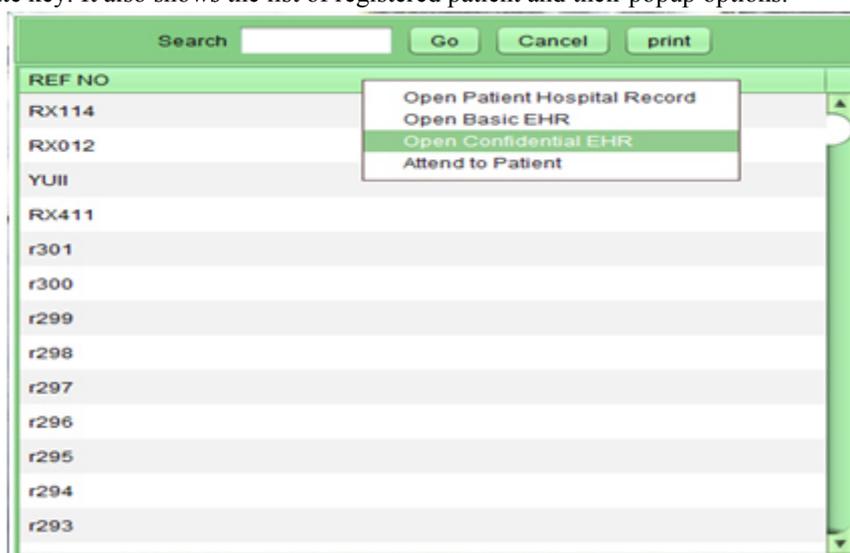

Figure 10 Medical Doctor's Interface

*4.1.9 Confidential EHR Access*
Corresponding operation are performed depending on the popup option chosen, Medical doctors who selected 'Open confidential EHR' can access a patient confidential record by using the patient private key. Figure 11 represents the interface whereby the medical doctor is required to enter the patient private key. The confidential EHR shows after the doctor has entered the patient private key. Figure 12 represents the confidential EHR

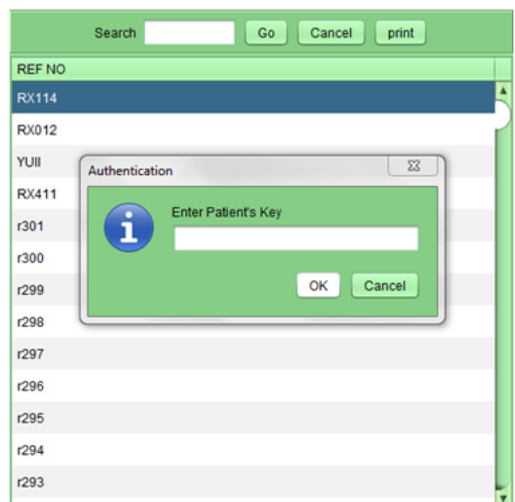

Figure 11 Confidential EHR Access

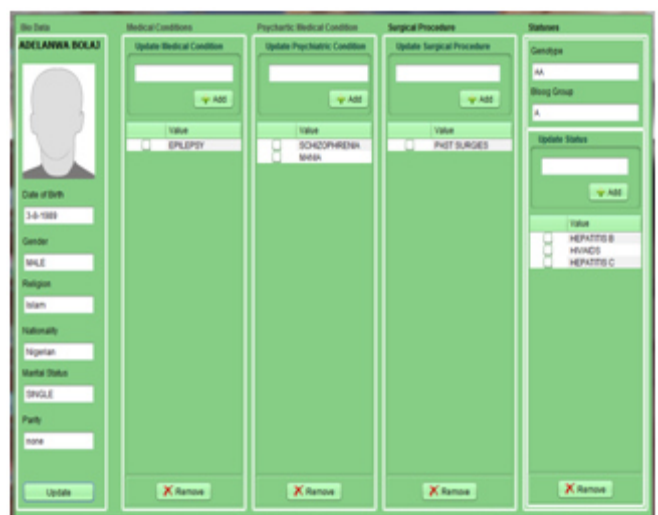

Figure 12 Confidential EHR

*4.2 System Evaluation*

 The system was evaluated comparing both iris biometrics templates and fingerprint biometric template with respect to their response time over 200 database of iris and fingerprints template. From Figure 13 it can be concluded that fingerprint biometric template has a higher response time compared to that of iris biometrics. Average response time for fingerprint is 6seconds and the average response time for iris is 11.1seconds. The larger the biometrics database the longer it takes to fetch record, fingerprints appeared to produce better result in terms of the response time than iris.

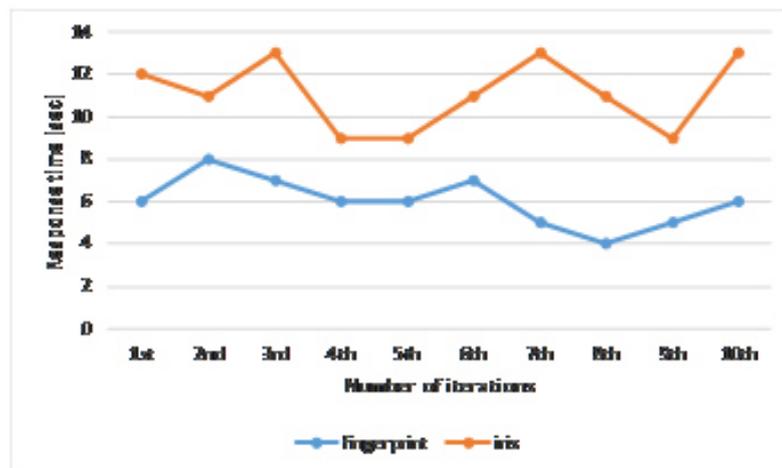

Figure 13 System Evaluation

## 5. Discussion

In conclusion, the system protects privacy and confidentiality by limiting the amount of data exposed to users. These limitations prevent attackers from gaining access to information that may harm both users and providers, yet the system was able to do its job by listing all the necessary data. The system also enables emergency medical technicians to gain easy and reliable access to necessary attributes of patients' EHR while still maintaining the privacy and confidentiality of the data using the patient's fingerprint and iris. Patients are also able to control who should have access to their confidential health record by giving their private keys only to doctors that should access their confidential attributes. Reliability is employed by exploiting the uniqueness of a person's fingerprint and iris as a means of access control as well as the precision of fingerprint scanners and iris scanners. The system also provides applicable security when patients' records are shared either with other practitioners, employers, organizations or research institutes. System evaluation was performed by comparing the response time of fingerprint authentication to that of iris authentication. The result of the evaluation shows that fingerprint has a higher response time with average of 6seconds when 200 patients biometric data are registered in the database.

Future work should make use of more questionnaires so that sufficient input can be gathered to work with and should also consider more diseases in the various categories. Further research should also include identification of system requirements in order to deploy this system with a larger biometrics dataset.

## 6. Acknowledgments